\documentclass[aoas]{imsart}

\RequirePackage{amsthm,amsmath,amsfonts,amssymb,mathabx}
\RequirePackage[authoryear]{natbib}
\RequirePackage[colorlinks,citecolor=blue,urlcolor=blue]{hyperref}
\RequirePackage{graphicx}
\RequirePackage{xcolor}
\RequirePackage{soul}

\usepackage{amssymb}
\usepackage{amsbsy}
\usepackage{amsmath}
\usepackage{color}
\usepackage{psfrag,epsfig, url}
\usepackage{array}

\startlocaldefs

\DeclareMathOperator{\arctanh}{arctanh}

\newif\ifnotesw \noteswtrue

\theoremstyle{plain}
\newtheorem{axiom}{Axiom}

\newtheorem{remark}[axiom]{Remark}
\theoremstyle{remark}

\endlocaldefs

\begin{document}

\begin{frontmatter}
\title{Mapping Incidence and Prevalence Peak Data for SIR Forecasting Applications}
\runtitle{Incorporating Incidence Peak and Time Data for SIR Forecasting Applications}

\begin{aug}
\author[A]{\fnms{Alexander C.}~\snm{Murph}\ead[label=e1]{murph@lanl.gov}\orcid{0000-0001-7170-867X} },
\author[A]{\fnms{G. Casey}~\snm{Gibson}\ead[label=e2]{gcgibson@lanl.gov}\orcid{0000-0002-0370-9846}},
\author[A]{\fnms{Lauren J.}~\snm{Beesley}\ead[label=e3]{lvandervort@lanl.gov}\orcid{0000-0002-3788-5944}},
\author[B,C]{\fnms{Nishant}~\snm{Panda}\ead[label=e4]{nishpan@lanl.gov}\orcid{0000-0001-9754-2794} },
\author[B]{\fnms{Lauren A.}~\snm{Castro}\ead[label=e5]{lcastro@lanl.gov}\orcid{0000-0002-9778-570X} },
\author[B]{\fnms{Sara}~\snm{Del Valle}\ead[label=e6]{sdelvall@lanl.gov}\orcid{0000-0002-0159-1952} },
\and
\author[A]{\fnms{Dave}~\snm{Osthus}\ead[label=e7]{dosthus@lanl.gov}\orcid{0000-0002-4681-091X} }

\address[A]{Statistical Sciences (CCS-6), Computer, Computational, and Statistical Sciences
Division, Los Alamos National Laboratory, Los Alamos, NM, 87545 \printead[presep={,\ }]{e1,e2,e3,e7}}
\address[B]{Data Analytics \& Forecasting, A-1: Information Systems \& Modeling, Los Alamos National Laboratory, Los Alamos, NM, 87545 \printead[presep={,\ }]{e5,e6}}
\address[C]{Information Sciences (CCS-3), Theoretical Division, Los Alamos National Laboratory, Los Alamos, NM, 87545 \printead[presep={,\ }]{e4}}
\end{aug}

\begin{abstract}
Infectious disease modeling and forecasting have played a key role in helping assess and respond to epidemics and pandemics. Recent work has leveraged data on disease peak infection and peak hospital incidence to fit compartmental models for the purpose of forecasting and describing the dynamics of a disease outbreak.  Incorporating these data can greatly stabilize a compartmental model fit on early observations, where slight perturbations in the data may lead to model fits that project wildly unrealistic peak infection.  We introduce a new method for incorporating historic data on the value and time of peak incidence of hospitalization into the fit for a Susceptible-Infectious-Recovered (SIR) model by formulating the relationship between an SIR model's starting parameters and peak incidence as a system of two equations that can be solved computationally.  This approach is assessed for practicality in terms of accuracy and speed of computation via simulation.  To exhibit the modeling potential, we update the Dirichlet-Beta State Space modeling framework to use hospital incidence data, as this framework was previously formulated to incorporate only data on total infections.
\end{abstract}


\end{frontmatter}

\section{Introduction}
Compartmental models have seen broad usage at the onset of several disease outbreaks in the last century as a means to project expected numbers of infected and to drive healthcare response.  Broadly speaking, a compartmental model describes the dynamics of a disease spread by breaking a population into set categories and modeling the process by which individuals transition through these categories.  Perhaps the most basic is the Kermack–McKendrick model, often called the SIR or Susceptible-Infectious-Recovered model, which models the movement of subjects from being Susceptible, to Infected (and contagious), and then to Removed \citep{kermack1927}.  Since the development of the SIR model, further research has extended the idea to include additional compartments, such as the Exposed category in the SEIR model -- where a subject has been exposed to the disease but is not yet infectious -- and the ability to move back to the Susceptible category in the SEIS model (see \citet{walter1999, hethcote2000} for an overview of each).  Compartmental models have been applied to projection tasks such as the 2014-15 Ebola epidemic \citep{chretien2015}, the 2009 A/H1N1 influenza pandemic \citep{nsoesie2014}, several HIV outbreaks \citep{anderson1988, golub1993, nyabadza2011}, the recent COVID-19 pandemic \citep{zhao2020, cooper2020, zhang2022}, and many other epidemiological projection tasks in the last century (see, for instance, \citet{guanghong2004, ladeau2011, zhan2019} and references therein).

The elegance of compartmental models is in their succinct ability to describe the state of a population in terms of how subjects transfer to and from the different categories, and thus fitting these models requires estimation of interpretable quantities such as rates of infection and recovery.  For instance, denote the proportion of the population in the susceptible, infectious, and removed compartments by $S_t, I_t,$ and $R_t,$ respectively, such that $S_t + I_t + R_t = 1$ for all $t$.  Then the SIR model is determined by the equations
\begin{subequations}
\begin{align}
    \frac{dS_t}{dt} &= - \beta S_t I_t \label{eq:sir_1}, \\ 
    \frac{dI_t}{dt} &= \beta S_t I_t - \gamma I_t \label{eq:sir_2}, \\
    \frac{dR_t}{dt} &= \gamma I_t \label{eq:sir_3} ,
\end{align}
\end{subequations}
where $\beta > 0$ is the disease transmission rate and $\gamma >0$ is the rate of recovery.  If one knew these two rates, and the initial number of individuals in each category -- $S_0, I_0$ and $R_0$ -- the numbers $S_t, I_t, \text{and} \; R_t$ could be numerically simulated for any time-point $t$.  While modern computational tools do make simulating these quantities feasible, the need to simulate the entire system numerically to get the exact counts in each compartment is challenging. That is, fitting these models to data and can be computationally expensive for more intricate compartmental structures.  Researchers have determined analytic solutions for the entire SIR model (explicit forms/approximations to the number of susceptible, infected, and removed at a specific time) that do not require numeric simulation \citep{harko2014,schlickeiser2021,carvalho2021}.  However, these solutions involve reparameterizing the time axis, and explicit calculations back onto the original time axis require numeric integration or approximation methods.  Similar approaches are also used to obtain exact and approximate solutions to the more complicated compartmental models, such as the SEIR model \citep{wang2014, piovella2020} and the SIRS model \citep{acedo2010}.

Analytic maps from the initial starting parameters of an SIR curve to quantities of interest (QoIs) -- such as the value of the peak of the infected curve and the limiting number of susceptible individuals -- were developed as early as the mid-1900s \citep{kendall1956, bailey1957, hethcote1976}.  The focus on these quantities has generally been motivated by their usefulness to making public health decisions.  For instance, in papers such as \cite{hethcote2000} and \cite{weiss2013}, the maximum value of the infected compartment (not the time of maximum infection) was studied for the purpose of informing public health officials of the maximum number of infections after the initial estimation of the disease transmission and recovery rates, since knowing this maximum quantity informs how many hospital beds might be needed.  The authors in \citet{castro2020} point out that the time of peak infection is also informative for healthcare officials, and they develop an analytic form for the peak time of infection using an approximation of the SIR curve.  An exact form for the peak infection time is available in \citet{kendall1956} and \citet{deakin1975}, albeit in terms of an integral without a closed form.  Several modern papers study fast approximations to the integral form for peak infection time \citep{cadoni2020, turkyilmazoglu2021}, using the previously-mentioned analytic solutions to the SIR model. These different approximations for peak infection time are evaluated and compared in \citet{kroger2021}.

Some recent papers study SIR curve QoIs for inferential tasks related to modeling an epidemic \citep{miller2009,lang2018}.  In \citet{amaro2023}, the Gumbel distribution is suggested as a good approximation of the infection curve in an SIR model, and maps between the SIR curve parameters and peak value/time are used to develop Method of Moments estimators of the Gumbel parameters.  The Gumbel distribution is then used to approximate the exact solution of the SIR model.  In \cite{osthus2017}, the relationship between peak infection value and the SIR parameters is used to incorporate historical data on epidemic peaks into the inferential task of fitting an SIR curve during the early stages of an epidemic. The authors point out that an SIR curve is sensitive to small perturbations in the transmission and recovery rates, and that incorporating these data discourages models that well-represent early data but drastically over-predict the peak infection value.  In \cite{mcandrew2024}, both historic and surveyed QoIs are used to constrain an SEIRH model (where the added ``H" category refers to people hospitalized at a given time) in much the same way as \cite{osthus2017}.

\begin{figure} 
\includegraphics[width=\textwidth]{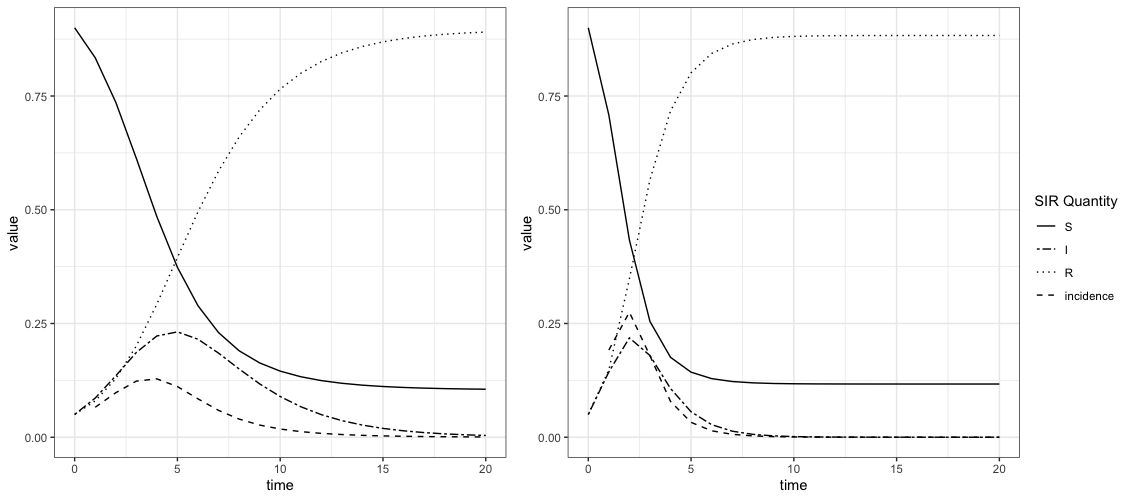}
\caption{Two SIR models with incidence and starting values $S_0 = 0.9$ and $I_0 = R_0 = 0.05$.  The plot on the left uses parameters $(\beta, \gamma) = (1.137, 0.446)$; the plot on the right uses $(2.592, 1.058)$. } \label{fig:sir_example}
\end{figure}

Both \citet{amaro2023} and \citet{osthus2017} use information on peak infection value and time to inform the modeling of a pandemic and epidemic, respectively. Since most modern monitoring systems approximate the daily number of \textit{new} cases of a disease, called \textit{incidence}, this is not the most practically useful development.  The number of infected individuals at any given time, called \textit{prevalence}, is typically an unobserved quantity \citep{noordzij2010}.  For examples of all quantities in an SIR curve on a fixed time axis, see Figure \ref{fig:sir_example}.  Note that peak prevalence and peak incidence need not occur at the same time, and that it is mathematically possible for the incidence to be greater than prevalence (as visualized on the right side of Figure \ref{fig:sir_example}).  This being said, in most disease outbreaks of note, prevalence is typically greater than incidence.

The model in \cite{mcandrew2024} incorporates incidence QoI data by extending the parameter space to include peak incidence value (PIV) and peak incidence time (PIT) and by defining a prior on these values using either historic or survey data.  While this model achieves a similar goal to the ultimate goal of this paper, its main distinction is that that it requires an importance sampling scheme to fit, which we outline here for completeness: priors on the rate parameters and initial values for the compartmental model are used to sample proposal values, the entire system is numerically simulated to determine the (PIV, PIT), then the likelihood of this system is determined via the prior probability of (PIV, PIT).  Note that while this need not be true for the SEIRH model considered in \cite{mcandrew2024} (since it is not an SIR model), defining both a prior on the rate parameters and on PIV and PIT for an SIR model is redundant given the initial values $S_0$, $I_0$, and $R_0$, since a given set of starting parameters should immediately determine these QoIs.  

Incorporating historical peak values and times of an outbreak into a Bayesian compartmental forecasting model via prior specifications is highly useful to constrain forecasts\footnote{For a review of the differences between disease forecasts and disease projections, see \cite{massad2005}.}, especially early in the outbreak (see \cite{osthus2017}, \cite{osthus2019}, and \cite{mcandrew2024}).  Relating these historical QoIs requires determining a relationship between historical peak values and times as well as the parameters and initial conditions of the compartmental model (e.g., the SIR model).  The challenge lies in this relationship. Analytically, this relationship is known for observed peak prevalence value (PPV) and peak prevalence time (PPT). However, almost all infectious disease data is of incidence. Despite being principally unsound, the data application in \cite{osthus2017} treated incidence data as if it were prevalence in order to use the known analytic relationships.  In this paper, we make three contributions.  First, we develop the methods to map peak values and times of \textit{incidence} to the parameters of the SIR model, given the initial conditions.  Second, we demonstrate how to incorporate these new mappings into the model of \cite{osthus2017} and provide the code used to do so.   Third, we compare the forecasts of the modeling framework of \cite{osthus2017} using both incidence data as incidence, and using the misspecified prevalence data.  In addition to comparing these forecasts, we show that if SIR parameter inference is desired, then mistaking incidence data for prevalence data will result in biased estimation.

This paper is outlined as follows.  In Section \ref{sec:existing_methods}, we review existing methods for mapping SIR parameters to peak/time of prevalence.  In Section \ref{sec:new_maps}, we develop identical maps for incidence, then provide methods for inverting these maps for both incidence and prevalence.  In Section \ref{sec:modeling_framework}, we develop and improve upon the modeling framework in \citet{osthus2017} to incorporate historic incidence peak/time data when fitting an SIR curve.  We apply this model to influenza data in Section \ref{sec:application}.

\section{Analytic Solutions to the SIR Equations for Incorporating Prevalence Data} \label{sec:existing_methods}
The system \eqref{eq:sir_1}-\eqref{eq:sir_3} can be solved analytically for all time by reparameterizing the time axis to be instead in terms of the number of individuals removed from the system beyond the initial amount in the Removed category $R_0$,
\begin{subequations}
    \begin{align}
        S_t &= S_0 e^{-\beta \tau_t} \label{eq:sir_analytic_1}, \\
        I_t &= S_0 + I_0 - S_0 e^{-\beta \tau_t} - \gamma \tau_t \label{eq:sir_analytic_2}, \\
        R_t &= R_0 + \gamma \tau_t, \label{eq:sir_analytic_3}
    \end{align}
\end{subequations}
where $\tau_t$ is the inverse of the map
\begin{equation} \label{eq:time_reparam}
    t_{\tau} := \int_{0}^\tau \left[ \frac{d\tau'}{S_0 + I_0 - S_0 e^{\beta \tau'} - \gamma \tau'} \right]^+,
\end{equation}
where $[x]^+ := \max\{x,0\}.$ The above form for the SIR dynamics is particularly useful since it allows one to calculate the number of individuals in each category without numerically simulating the entire system.  The major drawback of this form is that mapping values back to an interpretable time requires one to approximate the integral in \eqref{eq:time_reparam}, since this integral is nonelementary.  There exist other methods that derive an analytic solution to the SIR system \citep{harko2014}.  We use this formulation since it is well-known (having been originally developed in \cite{kendall1956}), and because it simplifies the calculation of SIR curve QoIs, such as PPV and PPT.  For instance, the maximum of \eqref{eq:sir_analytic_2} with respect to $\tau$ occurs at
\begin{equation} \label{eq:max_tau}
    \tau^* = \frac{1}{\beta} \log \left( \frac{\beta S_0}{\gamma} \right),
\end{equation}
which can be derived via straightforward calculus.  PPT can then be approximated from \eqref{eq:time_reparam}, and PPV calculated from \eqref{eq:sir_analytic_2}.

The process above of calculating PPV and PPT, denoted $(I_{t_{\tau^*}}, t_{\tau^*}),$ from the initial starting parameters can be reversed for fixed $S_0, I_0, R_0$.  That is, given $(I_{t_{\tau^*}}, t_{\tau^*})$, we propose a method to calculate $(\beta, \gamma)$.  Define $\rho$ as 
\begin{equation} \label{eq:rho}
\rho = \frac{\beta}{\gamma}.
\end{equation}
The quantity $S_0 \rho$ is more widely known as the \textit{basic reproduction number}, which measures the average number of additional infections generated by a single new infection \citep{cadoni2020}.  Plugging \eqref{eq:max_tau} into \eqref{eq:sir_analytic_2} gives
\begin{equation}
    I_t = S_0 + I_0 - \frac{1}{\rho} - \frac{\log (\rho S_0)}{\rho}.
\end{equation}
From here, setting $I_t:= I_{t_{\tau^*}}$ and solving for $\rho$ is a (piecewise) convex optimization problem, which is typically fast and accurate computationally.  Note that a change of variables for \eqref{eq:time_reparam} gives
\begin{equation} \label{eq:time_reparam_changeofvars}
    \beta t_{\tau} = \int_{0}^{\beta \tau} \left[ \frac{d\hat{\tau} }{S_0 + I_0 - S_0 e^{-\hat{\tau}} - \hat{\tau} / \rho }\right]^+ ,
\end{equation}
and that $\beta \tau^* = \log(\rho S_0)$.  Thus, once the value of $\rho$ is approximated, \eqref{eq:time_reparam_changeofvars} can be evaluated to get $\beta t_{\tau^*}$.  Since $t_{\tau^*}$ is assumed known \textit{a priori}, the values of $\beta$ and $\gamma$ can be obtained via arithmetic.  

We introduce the above derivations not only because they will be used in the next Section, but also to motivate the type of calculation we aim to develop in this paper.  While the above approach still requires computational methods to map from PPV and PPT to the SIR model parameters, this approach is much more direct than the brute force method of simulating several $(\beta, \gamma)$ combinations until a prevalence curve with a peak sufficiently near $(I_{t_{\tau^*}}, t_{\tau^*})$ is discovered \citep{prangle2016}. 

\section{Mapping Incidence Data to SIR Parameters}\label{sec:new_maps}

As mentioned above, a primary focus of this work will be to develop maps between PIV and PIT and the initial SIR curve parameters.  In this direction, one can derive an equivalent formulation of the SIR dynamics in \eqref{eq:sir_1} - \eqref{eq:sir_3} by replacing \eqref{eq:sir_2} with
\begin{align} 
    I_t &= I_{t-1} + \beta S_{t-1} I_{t-1} - \gamma I_{t-1} \label{eq:prevalence_w_incidence}.
\end{align}
Putting \eqref{eq:prevalence_w_incidence} into words, prevalence is equal to the prevalence at the last time step, plus those in the infected category that infect those in the susceptible category, minus the number in the infected category that are removed from the system.  The form for prevalence in \eqref{eq:prevalence_w_incidence} is particularly useful for this application since it contains a term that explicitly models incidence at time $t$: $\beta S_{t-1} I_{t-1}$.

\subsection{Mapping SIR Parameters to Peak Incidence Value and Time} \label{subsec:sir_to_incidence}
Reparameterizing the time axis for the term $\beta S_{t} I_t$ using \eqref{eq:sir_analytic_1} and \eqref{eq:sir_analytic_2} gives the following form
\begin{equation} \label{eq:incidence_analytic}
    \beta (S_0 e^{-\beta \tau_t}) (S_0 + I_0 - S_0 e^{-\beta \tau_t} - \gamma \tau_t ).
\end{equation}
The value for $\tau$ that maximizes \eqref{eq:incidence_analytic} also satisfies the equation,
\begin{equation} \label{eq:incidence_firstderiv}
    -(S_0 + I_0) + \gamma \tau + 2S_0 e^{-\beta \tau} - \frac{1}{\rho} = 0.
\end{equation}
Solving \eqref{eq:incidence_firstderiv} is also a convex optimization problem on a single parameter, and is thus feasible and accurate to do numerically.  A solution to \eqref{eq:incidence_firstderiv} gives the $\tau$ for the timepoint directly \textit{before} the time of max incidence, $\tau_{t^*-1}$, where $t^*$ denotes the timepoint of max incidence.  Of course, this is all that is required to calculate the PIV, $\beta S_{t-1} I_{t-1},$ by a direct application of \eqref{eq:incidence_analytic}.  The calculation of PIT is similarly straightforward.  Using \eqref{eq:sir_analytic_2}, the prevalence at $\tau_{t^*-1}$ can be directly calculated.  The prevalence value for $\tau_{t^*-1}$ and \eqref{eq:sir_analytic_3} then gives $\tau_{t^*},$ and \eqref{eq:time_reparam} can then be used to calculate $t^*$,
\[ t^* = \int_{0}^{\tau_{t^*}} \left[ \frac{d\tau'}{S_0 + I_0 - S_0 e^{\beta \tau'} - \gamma \tau'} \right]^+. \]

\subsection{Mapping Incidence Peak Value and Time to SIR Parameters} \label{subsec:sir_to_incidence_map}

The proposed method to map PIV and PIT back to the parameters of an SIR curve solves the following system of equations implied by \eqref{eq:incidence_analytic} and \eqref{eq:incidence_firstderiv}:
\begin{align}
    \beta (S_0 e^{-\beta \tau^\star }) (S_0 + I_0 - S_0 e^{-\beta \tau^\star} - \gamma \tau^\star ) &= \text{PIV} \label{eq:system_of_eqns_1} \\
    -(S_0 + I_0) + \gamma \tau^\star + 2S_0 e^{-\beta \tau} - \frac{1}{\rho} &= 0 \label{eq:system_of_eqns_2},
\end{align}
where $\tau^\star$ solves the equation
\begin{equation} \label{eq:time_integral_solve}
    \int_{0}^{\tau^\star} \left[ \frac{d\tau'}{S_0 + I_0 - S_0 e^{\beta \tau'} - \gamma \tau'} \right]^+ = \text{PIT}.
\end{equation}

While solving a system of two equations with two unknowns (for $\beta$ and $\gamma$) is generally feasible computationally, the major bottleneck for this problem is the need to invert \eqref{eq:time_integral_solve}.  A brute-force computational approach to solving this system of equations would require one to both invert and solve \eqref{eq:time_integral_solve} for every value of $(\beta, \gamma)$ investigated.  While experiments in this direction have proven to be surprisingly fast and accurate, the confounding computational approximations encourage a more analytic solution, or alternative computational strategies.  We consider three possible alternatives to computationally estimating the integral (referred to the ``Compute Integral" method in the subsequent).

\subsubsection{Taylor Approximation} \label{subsubsec:taylor}
Given a candidate $(\beta, \gamma)$ in any numerical solver, we approximate $\tau^\star$ by taking the second degree Taylor expansion of the integral in \eqref{eq:time_integral_solve} and solving for $\tau^\star$ algebraically \citep{murray2002}.  This leads to the following closed-form approximation,
\begin{equation}
    \tau^\star = \frac{\beta^2}{S_0} \left[ \left( \rho S_0 - 1 \right) + \kappa \tanh \left( \frac{\gamma \kappa (\text{PIT})}{2} - \phi \right)\right] + R_0,
\end{equation}
where
\begin{align*}
    \kappa &= \sqrt{\left( S_0 \rho - 1 \right)^2 + 2 S_0 I_0 \rho^2 }, \\
    \phi &= \frac{1}{\kappa} \arctanh \left[ S_0 \rho - 1 \right].
\end{align*}
Using this approximation for $\tau^\star$, we numerically solve the system of equations expressed by \eqref{eq:system_of_eqns_1} and \eqref{eq:system_of_eqns_2}.

\subsubsection{Single ODE Approximation} \label{subsubsec:partialode}
Instead of using a Taylor approximation to estimate $\tau^\star$ it is possible to do so by numerically solving an Ordinary Differential Equation (ODE).  Using the definition of $\tau$, we combine equations \eqref{eq:sir_analytic_1} and \eqref{eq:sir_analytic_3} to get
\begin{equation}
    S_t = S_0 \exp \left( -\rho ( R_t - R_0 ) \right).
\end{equation}
From here, we use \eqref{eq:sir_3} and the assumption that $I_t = 1 - S_t - R_t$ to get the ODE,
\begin{equation} \label{eq:partialODE}
    \frac{dR_t}{dt} = \gamma \left[ 1 - S_0 e^{-\rho(R_t - R_0)} - R_t \right].
\end{equation}
The authors in \cite{cadoni2020} point out that while \eqref{eq:partialODE} is a transcendental equation, it can be solved numerically, and has a single, unique solution by the general existence and uniqueness theorem for the solutions of ODEs.  Using this approximation for $\tau^\star$, we numerically solve the system of equations expressed by \eqref{eq:system_of_eqns_1} and \eqref{eq:system_of_eqns_2}.

\subsubsection{Full ODE Approximation} \label{subsubsec:fullode}
As a final computational method for mapping PIV and PIT to $\beta$ and $\gamma$, we numerically solve the system of ODEs described in \eqref{eq:sir_1} - \eqref{eq:sir_3}.  In an optimization algorithm, this would require the ODE to be solved for every possible $(\beta, \gamma)$ pair queried.  Much like the brute force computational method, we expect this method to be accurate, but to come at a high computational cost.

\subsubsection{Comparision via Simulation} \label{subsubsec:sim_study} We compare all the above approaches via simulation, repeating the following steps 1000 times:
\begin{enumerate}
    \item Sample a (PPV, PPT) from the bivariate normal $\mathcal{N}(\mu, \Sigma)$ where $\mu = (0.0144, 17.9)$,
    \[ \Sigma = \begin{pmatrix}
 0.000036 & 0.0187 \\
-0.0187 & 16.09
    \end{pmatrix}, \]
    truncated so that PPV $\in (\theta_{I_0} , 1)$ and PPT $\in (1,35)$.  This corresponds to the set of feasible values and sampling distribution described in \cite{osthus2017};
    \item Use the method from Section \ref{sec:existing_methods} to map this back to $(\beta, \gamma)$, then numerically simulate the system in \eqref{eq:sir_1}-\eqref{eq:sir_3} to get the ``true" values for PIT and PIV;
    \item With PIT and PIV, use each method described above to find approximate values $(\hat{\beta}, \hat{\gamma})$;
    \item For all approximation methods, compare the estimated $\widehat{\text{PIT}}$ and $\widehat{\text{PIV}}$ (gotten by numerically simulating the system from the appropriate $(\hat{\beta}, \hat{\gamma})$) against the true PIT and PIV.
\end{enumerate}
We outline the results from the above simulation in Table \ref{tab:sim}.  While the fastest method is the Taylor approximation, this method is also the least accurate. This is as expected, since this approximation is only accurate for sufficiently small values of $\rho (R_t - R_0)$ \citep{murray2002}.  The Single ODE approximation method is slightly more accurate than the Taylor approximation, but at a higher computational cost.  The Compute Integral and the Full ODE approximation are comparable, with the Compute Integral method being more accurate and the Full ODE method being faster.

While the Compute Integral method is the slowest of these approaches, it is still somewhat fast (taking around half a second on average), and it is the most accurate overall (since the Taylor approximation and Single ODE methods are greatly off for PIV).  Since the data sizes for PIT and PIV data are not exorbitantly large, this computational burden would be acceptable in converting a data set of $(\text{PIV}, \text{PIT})$ values to $(\hat{\beta}, \hat{\gamma})$ values.  For the applications in this paper, we will use the Compute Integral method. Future work might examine alternative approximation methods, especially for approximating the inverse of the integral in \eqref{eq:time_integral_solve}.  

\begin{table}
\begin{center}
{\setlength{\extrarowheight}{2pt}%
\begin{tabular}{||c | c c c c||} 
 \hline
 Quantity & Compute Integral & Taylor Approx. & Single ODE & Full ODE  \\ [0.5ex] 
 \hline\hline
 Avg. PIV Error & 4.07e$^{-4}$ & 1.51e$^{-4}$ & 1.36e$^{-4}$ & 16.02e$^{-4}$ \\ 
 \hline
 Std. Dev. PIV Error & 1.836e$^{-4}$ & 0.566e$^{-4}$ & 0.660e$^{-4}$ & 6.848e$^{-4}$ \\
 \hline
  Avg. PIT Error & 0 & 3.209 & 2.236 & 0 \\ 
 \hline
 Std. Dev. PIT Error & 0 & 1.888 & 1.647 & 0 \\
 \hline
 Avg. Runtime & 5.626e$^{-1}$ & 0.035e$^{-1}$ & 2.017e$^{-1}$ & 0.605e$^{-1}$ \\
 \hline
 Std. Dev. Runtime & 2.060e$^{-1}$ & 0.015e$^{-1}$ & 0.510e$^{-1}$ & 0.324e$^{-1}$ \\ 
 \hline
\end{tabular}}
\end{center}\caption{Errors and runtimes (in seconds) for the four approximation methods described in Section \ref{subsec:sir_to_incidence_map}, using the simulation study described in Section \ref{subsubsec:sim_study}.  The methods are the Compute Integral approach (top of \ref{subsec:sir_to_incidence_map}), the Taylor approximation (\ref{subsubsec:taylor}), the single ODE approximation (\ref{subsubsec:partialode}), and the full ODE approximation (\ref{subsubsec:fullode}). }\label{tab:sim}
\end{table}

\section{A Bayesian State-Space SIR Model} \label{sec:modeling_framework}
In this section, we introduce the Dirichlet-Beta state-space model (DBSSM) from \cite{osthus2017} and update it to incorporate historic PIV and PIT data.  The original formulation of the DBSSM was to answer an observed issue associated with using the SIR model for early-pandemic forecasting tasks.  Namely, that two SIR curves that reasonably fit early count data may lead to drastically different PPV predictions.  In a simulation example, \cite{osthus2017} show two such SIR curves that have peaks that differ by 30\% of the entire population, even though they have a nearly-identical fit to the early-pandemic data observations (see Figure 3 in the cited paper).  To address this stability issue, the DBSSM incorporates historic PPV and PPT data into the prior specifications to discourage SIR curve fits with peak values that are greatly above reasonable expectations.  As mentioned previously, this incorporation of prevalence data is not the most practical approach, since incidence data is generally the observed quantity.  A further shortcoming in the original DBSSM formulation is that it learns a map between PPT and the SIR curve parameters, rather than using an analytic map.  After introducing this model, we will identify ways that the methodology developed in this paper will improve these issues for the DBSSM.

Let $y_t$ be the observed proportion in a population that tested positive for some disease at some timepoint $t$, and let $\theta_t = (S_t, I_t, R_t)'$. Then the DBSSM is defined as,
\begin{equation} \label{eq:dbssm_data}
    y_t | \theta_t, \phi \sim \text{Beta}\left(\lambda \text{In}_t, \lambda (1 - \text{In}_t) \right)
\end{equation}
\begin{equation} \label{eq:dbssm_dirichlet}
    \theta_t | \theta_{t-1}, \phi \sim \text{Dirichlet}\left(\iota f(\theta_{t-1}, \beta, \gamma) \right),
\end{equation}
where $\phi = \{S_0, I_0, R_0, \beta, \gamma, \lambda, \iota\}$, $\lambda, \iota$ are variance control parameters, and $f$ is a map that propagates the SIR system determined by $(\theta_{t-1}, \beta, \gamma)$ forward one step according to \eqref{eq:sir_1}-\eqref{eq:sir_3}.  Note that, by this set up, the set of parameters $\theta_{0:t'} = \{ \theta_0, \theta_1, \dots, \theta_{t'} \}$ is a first-order Markov chain, and that for all $s\neq t$, the data observations $y_s$ and $y_t$ are independent given $\theta_t$.  The variable $\text{In}_t$ denotes the \textit{incidence} of the system at time $t$; in the original formulation of this model, the prevalence -- $I_t$ -- was used here.  The incidence at time $t$ is directly calculated using $\theta_t$ and \eqref{eq:prevalence_w_incidence}.

The conditional expectations of the model described by \eqref{eq:dbssm_data} and \eqref{eq:dbssm_dirichlet} are unbiased,
\begin{equation} \label{eq:dbssm_data_mean}
    E(y_t | \theta_t, \phi) = \text{In}_t 
\end{equation}
\begin{equation} \label{eq:dbssm_dirichlet_mean}
    E(\theta_t | \theta_{t-1}, \phi) = f(\theta_{t-1}, \beta, \gamma)
\end{equation}
while their respective variances reduce to zero as $\lambda, \iota \to \infty$.  Of course, the conditional mean in \eqref{eq:dbssm_dirichlet_mean} is dependant upon the accuracy of $f$ in propogating the latent space $\theta_{t-1}$ forward one time step.  The authors in \cite{osthus2017} used a fourth-order Runge-Kutta approximation and observed reasonable accuracy. 

We review the full Bayesian framework of the DBSSM and provide the prior specification in Appendix \ref{a:DBSSM}.  The main innovation of this model is that the parameter space is expanded to include the latent variable $z = (PPT, PPV)$, and this latent variable is given a prior that incorporates historic PPT and PPV data.  In the following, we review this prior, $\pi(z|\theta_0)$, and the mechanism by which this prior informs the SIR parameters $\beta$ and $\gamma$.  These priors are then each updated according to the theory developed in this paper.

\subsection{Specification of \texorpdfstring{$\pi(z | \theta_0)$}{TEXT}} The prior on $\pi(z | \theta_0)$ in the DBSSM is a minimal-assumption distribution on historic data on PPT and PPV.  Note that this is the mechanism by which the authors in \cite{osthus2017} directly address the aforementioned stability issues with fitting an SIR curve with early pandemic data.  They do so by fitting a normal distribution to historic influenza QoI data, truncated to enforce that an epidemic will occur (the lower bound on $PPV$ was set to $I_0$) and so that the peak happens within the influenza forecasting season ($PPT$ was required to be between the $1^{st}$ and $35^{th}$ weeks).  While somewhat loose, this prior gives very small (or zero) probability to values of $(PPV, PPT)$ that are drastically outside of historically observed pandemics.  

In the formulation of the DBSSM developed in this paper, the same prior used on PPT and PPV is now used on PIT and PIV.  To connect this constraint into the model, we must next define how the assumption on this latent space affects the learning of the SIR parameters $\beta$ and $\gamma$.

\subsection{Specification of \texorpdfstring{$\pi(\beta, \gamma | z, \theta_0)$}{TEXT}} 


With the addition of the latent variable $z$, the prior needed for the SIR parameters is $\pi(\beta, \gamma | z, \theta_0)$.  In the original formulation of the DBSSM, this prior is reparameterized according to $(\rho, \gamma)$, then factorized.  Thus, priors are instead given to $\pi(\rho | z, \theta_0)$ and $ \pi(\gamma | \rho, z, \theta_0)$.  This additional formulation is done to utilize the following analytic relationship from \cite{weiss2013},
\begin{equation}
    PPV = g_1(S_0, I_0, \rho) = I_0 + S_0 - S_0 \rho \left[ \log(S_0) + 1 - \log(S_0 \rho) \right].
\end{equation}
By inverting this relationship, samples from the latent quantity $z$ immediately determine the corresponding value of $\rho$.  The appropriate prior on this quantity would then be
\begin{equation} \label{eq:rho_prior_dave}
    \pi(\rho | z, \theta_0) ~\propto~ \delta (S_0 \rho - g_1^{-1}(PPV, S_0, I_0)),
\end{equation}
where $\delta$ is the Dirac delta function.  For the prior on $\gamma$, the map between $\gamma$ and PPT is estimated using a simulated data set of 5250 SIR curves.  This map,
\begin{equation}
    PPT = g_2(S_0, I_0, \rho, \gamma),
\end{equation}
was then used in lieu of any analytic form, and the prior on $\gamma$ was set to
\begin{equation} \label{eq:rho_prior_dave2}
    \pi(\rho | z, \theta_0) ~\propto~ \delta (S_0 \rho - g_2^{-1}(PPV, S_0, I_0, \rho)),
\end{equation}

We reiterate that there are two major shortcomings to the above prior specifications.  First, the above priors assume that there is access to historical PPV and PPT data, which is typically not the case, as public health data are generally on incidence, not prevalence.  In the original paper, incidence data were used instead of prevalence data without explicit justification.  Second, a map between PPT and the SIR parameters is estimated even though an analytic map between these quantities exists (see \eqref{eq:time_reparam} and \eqref{eq:max_tau}), unnecessarily introducing a source of uncertainty.

The methods developed in this paper correct the limitations found in the original formulation of the DBSSM, since they provide maps to replace $g_1, g_2$ above with
\begin{equation}
    (PIV, PIT) = h(S_0, I_0, \beta, \gamma),
\end{equation}
where $h^{-1}$ denotes the algorithm discussed in Section \ref{subsec:sir_to_incidence_map}.  Thus, the joint prior used for $\beta, \gamma$ in this updated version of the DBSSM is
\begin{equation} \label{eq:rho_prior_new}
    \pi(\beta, \gamma | z, \theta_0) ~\propto~ \delta \left( || (\beta, \gamma) - h^{-1}(PPV, S_0, I_0)||_1 \right),
\end{equation}
where $||\cdot ||_1$ is the 1-norm.

The naive treatment of incidence data as prevalence data (as was done in \cite{osthus2017}), need not necessarily lead to a loss of forecasting accuracy in the final model.  An incidence curve can well approximate, or even be equivalent to, a prevalence curve.  For instance, consider the SIR model constrained so that $\gamma = 1$, which corresponds to the Reed-Frost model \citep{abbey1952}.  In this case, incidence is precisely equal to prevalence, and thus either method for incorporating historic QoI data should yield an equivalent model.  This insight leads to the following Remark:

\begin{remark} \label{remark}
Using incidence QoI data in place of prevalence QoI data naively leads to an SIR curve where the $I$ compartment -- which normally corresponds to prevalence -- now models the progression of incidence.  Using historic incidence QoI data in the way outlined in this manuscript uses the incidence curve to model incidence as is desired.  While either can be viable for the purposes of \textbf{forecasting}, only the method described in this paper leads to \textbf{rate parameter estimates} that can be interpreted as infection and recovery rates. 
\end{remark}

\section{Application to Seasonal Influenza Data} \label{sec:application}
We recreate the data application from \cite{osthus2017}, using the updated model from Section \ref{sec:modeling_framework}.  The aim of this application is not to improve the forecasting in the original formulation of the DBSSM (see Remark \ref{remark}).  Rather, we will demonstrate that the forecasting capabilities of this model remain the same, while we also observe different estimates for the infection rate $\beta$, the recovery rate $\gamma$ and the basic reproduction number $\rho$.

The source data modified and then used for this application are counts of patients seen in the US with an influenza-like illness (ILI), where ILI is defined as having a temperature of at least 100 degrees Fahrenheit, a cough and/or a sore throat, and no known cause for those symptoms other than influenza \citep{CDC2024}.  These data are collected weekly, where more than 3400 outpatient healthcare providers report to the CDC the number of patients with ILI they treated \citep{CDC2023}.

The number of patients reported as having ILI will naturally also include cases of respiratory illnesses other than influenza.  Following the approach of \cite{shaman2013real}, we use virologic surveillance data (where patients are actually tested for influenza) to estimate the proportion of ILI patients with influenza, then multiply ILI data by this proportion.  This corrected data is referred to as ILI+.  For more details on this adjustment, see \cite{shaman2013real}. Note that the ILI+ data estimates the weekly incidence of influenza cases -- not prevalence.

\begin{figure} 
\includegraphics[width=\textwidth]{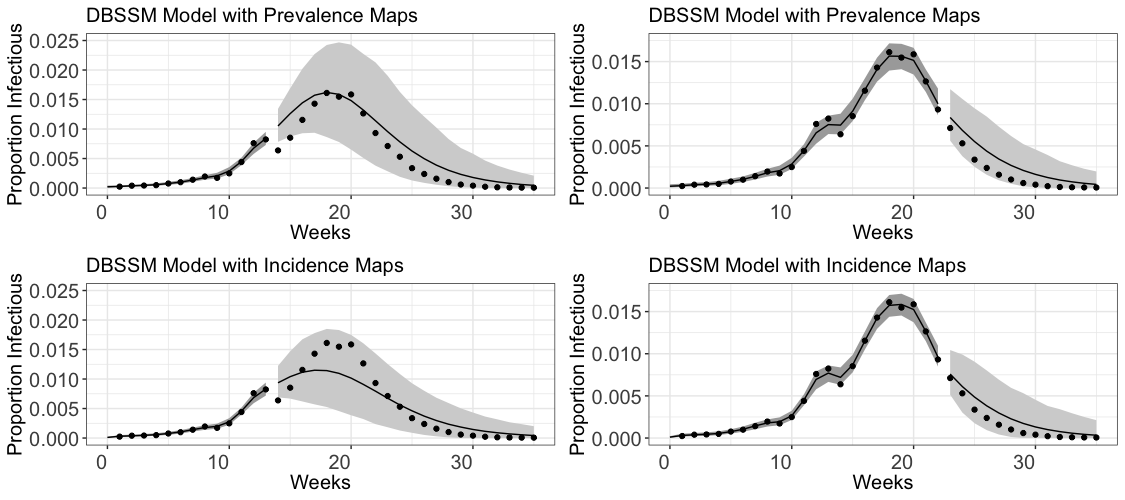}
\caption{The DBSSM fit to the 2010 US nationwide influenza outbreak starting at week 13 (left column) and week 22 (right column).  The top plots reproduces the forecasting model of \cite{osthus2017}, which incorrectly treated incidence data as though it were prevalence data. The bottom plots fit the forecasting model described in Section 3, which correctly treats incidence data as incidence data. Earlier on in the outbreak, the model developed in this paper has a tighter prediction interval.  The forecasts starting at week 22 are quite similar.} \label{fig:forecasting}
\end{figure}

To fit both versions of the DBSSM, we use ten influenza seasons: the seasons that started in the years 2002-2007, and the seasons the started during 2010-2013.  The years 2008 and 2009 were omitted to be consistent with \cite{osthus2017}; these two years correspond to a pandemic and the focus of that work was to forecast seasonal influenza.  Each season is defined as 35 consecutive weeks starting on roughly the first week of October (epidemiology week 40, treated as $t = 1$).

For a estimated proportion of individuals in a population infected with influenza at timepoints $\mathbb{T} = \{ 1, \dots, T\}$, suppose only the ILI+ data up through $t' \in \mathbb{T}$ are observed.  Given this, we simulate 62500 from the posterior $\pi(\boldsymbol{\theta_{{1:t'}}}, \boldsymbol{\phi} | y_{1:t'} )$ for four separate chains, discarding the first 12500 as burn-in and thinning out all but every tenth observation in the remaining samples.  Given these draws from the posterior distribution, the posterior predictive density, $\pi(y_{(t'+1):T} | y_{1:t'})$, is used to estimate ``future" observations of ILI+ data.  

We perform two separate fits of this posterior model, on the first thirteen days ($t'=13$) and on the first twenty two days ($t' = 22$), for the considered ILI+ data for the 2010 influenza season in the United States.  These fits are performed both using the original formulation of the DBSSM, which naively uses incidence data directly in place of prevalence data, and using the new formulation developed in this paper, which uses the new maps developed in Section \ref{sec:new_maps} for a more principled treatment of incidence data.  These fits and forecasts are outlined in Figure \ref{fig:forecasting}.  The dark shaded grey regions prior to $t'$ mark the 95 percentiles of the posterior density, while the lighter grey shaded regions after $t'$ make the 95\% prediction intervals.  The forecast using incidence data up until $t'=13$ has a narrower prediction interval than the one using prevalence data; each of the forecasts that use data up through $t'=22$ are laregly comparable.

\begin{figure} 
\hspace{-1.5cm}\begin{tabular}{|m{10em}|m{7em} m{7em} m{7em} |}
\hline
    Parameter & $\beta$ & $\gamma$ & $S_0 \beta / \gamma$ \\
    \hline \hline
    Prevalence Median & 2.15 & 1.60 & 1.21 \\
    Incidence Median & 3.21 & 2.66 & 1.09 \\
    \hline
\end{tabular}
\includegraphics[width=\textwidth]{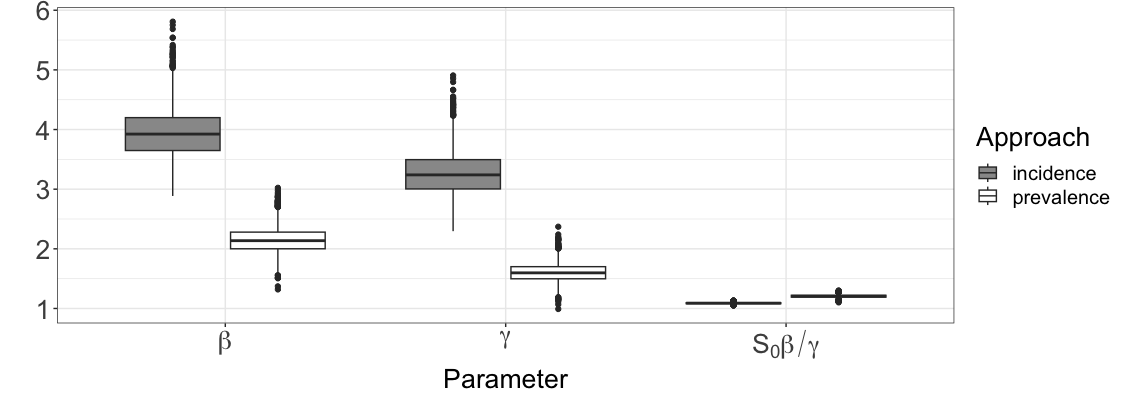}
\caption{Infection rates, recovery rates, and reproductions numbers drawn from the Gibbs sampler used to fit the DBSSM using both specifications of incidence data.  Using incidence data specified as prevalence leads to different estimates for these parameters.  These values were calculated using the data up through timepoint 22 ($t'=22$). } \label{fig:parameters}
\end{figure}

In addition to the slight improvements on forecasting we observe in Figure \ref{fig:forecasting}, this method also has strong implications for the interpretability of $\beta, \gamma$ for the fitted model.  Indeed, only the updated version of the DBSSM developed in this paper leads to realizations of these parameters that can accurately be interpreted as the infection rate ($\beta$), recovery rate ($\gamma$), and the basic reproduction number ($S_0 \rho$).  

\section{Discussion}\label{sec:discussion}
The main contribution of this paper is the development of methods to map the time and value of peak incidence to the SIR curve parameters, and vice versa, for the purpose of forecasting tasks and inference on disease rate parameters.  We do this by computationally solving a system of equations (\eqref{eq:incidence_analytic} and \eqref{eq:incidence_firstderiv}).  There are several impactful uses of these maps in the context of previous literature.  First, much like how the peak prevalence value (PPV) and time (PPT) are useful for public health response to an epidemic \citep{weiss2013}, the analogous quantities for incidence are also useful, since they describe the influx new patients entering the hospital system on a given day.  Second, this work improves upon existing work that uses historical prevalence data to model epidemics by creating a map from PIT and PIV to the SIR parameters, since incidence is typically the data that is available for ongoing epidemics \citep{osthus2017, amaro2023}.  In the case of the application in \cite{osthus2017}, where incidence data were used in place of prevalence data without justification, we have shown that this leads to biased SIR parameter estimates (see Figure \ref{fig:parameters}).  Furthermore, our results indicate that forecasts performed using the erroneous data specification leads to larger prediction intervals earlier on in the outbreak, although this forecast is largely comparable for the correct specification later on in the outbreak (see Figure \ref{fig:forecasting} and Remark \ref{remark}).  Of course, it remains more correct in principle to use incidence data appropriately when fitting compartment models for forecasting with ongoing incidence data \citep{nsoesie2013, chowell2016, abolmaali2021}.  We have provided a modeling framework that incorporates this data appropriately (see Section \ref{sec:modeling_framework}).   Lastly, while the methods discussed in \cite{mcandrew2024} do incorporate incidence data to fit forecasting models, this paper uses a Bayesian framework and importance sampling.  Since the maps developed here are deterministic, they can be used in both a Bayesian and a Frequentist framework.

As a direction for future work, it would be useful to investigate better approximations for the solution to \eqref{eq:time_integral_solve}.  The Taylor Approximation in Section \ref{subsubsec:taylor} was by far the fastest computationally, but it came with the highest error on PIT.  Finding a fast and accurate approximation to this equation would greatly increase the runtime for applications where the map between the SIR parameters and PIT/PIV must be evaluated several hundreds of thousands of times.  However, for most applications (including the one in this paper), the Compute Integral approximation is sufficiently fast.

As a second direction for future work, it would be interesting to investigate the analogous maps for more complicated compartmental models, such as the SEIR model, the SIRS model, and the SEIRH model.

\section*{Open Research Section}
All software used to perform the simulations and studies in this paper are publicly available at \url{https://github.com/lanl/precog}.  

\section*{Acknowledgments}
Research presented in this article was partially supported by the Laboratory Directed Research and Development program of Los Alamos National Laboratory under project number 20240066DR. Los Alamos National Laboratory is operated by Triad National Security, LLC, for the National Nuclear Security Administration of U.S. Department of Energy (Contract No. 89233218CNA000001).

This research was partially funded by NIH/NIGMS under grant R01GM130668-01 awarded to Sara Y. Del Valle.

\section*{Conflict of Interest}

The authors have no non-financial or other financial competing interests to declare that are relevant to the content of this article other than the aforementioned declared funding sources.

\bibliographystyle{imsart-nameyear} 
\bibliography{bibliography}       

\begin{appendix}
\section{Specification of the Dirichlet-Beta State-Space Model}\label{a:DBSSM}

The DBSSM is fit using a fully Bayesian framework, where the interest lies in estimating the joint posterior density of the latent space $\theta_{1:t'}$ and $\phi$ given the observed data $y_{1:t'}$.  Using the conditional independence assumption described above, this density can be written as follows:
\begin{equation}
    \pi(\theta_{1:t'}, \phi | y_{1:t'} )~ \propto ~ \pi(\phi) \pi(y_{1:t'}, \theta_{1:t'} | \phi) = \pi(\phi) \prod_{t = 1}^{t'} \mathcal{L} (y_t | \theta_t, \phi) \pi(\theta_t | \theta_{t-1}, \phi),
\end{equation}
where $\pi(\phi)$ is some prior on $\phi$, $\mathcal{L} (y_t | \theta_t, \phi)$ is the data likelihood determined by \eqref{eq:dbssm_data}, and the distribution $\pi(\theta_t | \theta_{t-1}, \phi)$ is determined by \eqref{eq:dbssm_dirichlet}.  To perform forecasts on observations $y_{t' : T}$, where $T$ is the final timepoint of the outbreak, one uses the posterior predictive distribution, where the model and latent-space parameters are integrated out:
\begin{equation}\label{eq:posterior_predictive}
    \pi(y_{ (t'+1):T} | y_{1 : t'} ) = \int \int \pi(y_{ (t'+1):T}, \theta_{1:T}, \phi | y_{1 : t'} ) d\theta_{1:T}, d\phi.
\end{equation}

To complete the specification of the DBSSM, one must determine what priors to put on the model parameters in $\phi$.  It is here that the authors in \cite{osthus2017} directly address the aforementioned stability issues with fitting an SIR curve with early pandemic data.  Define the latent variable $z = (PPT, PPV)$, and let $\boldsymbol{\phi}$ be the expanded model parameter vector that includes $z$.  We factorize the new prior distribution on $\boldsymbol{\phi}$ to get
\begin{equation}
    \pi(\boldsymbol{\phi}) = \pi(\iota) \pi(\lambda | \iota) \pi(\theta_0 | \lambda, \iota) \pi(z | \theta_0, \lambda, \iota) \pi(\beta, \gamma | z, \theta_0, \lambda, \iota).
\end{equation}
Several modeling assumptions on this conditional distribution give the abbreviated form,
\begin{equation} \label{eq:final_prior_form_new}
    \pi(\boldsymbol{\phi}) = \pi(\iota) \pi(\lambda ) \pi(\theta_0) \pi(z | \theta_0) \pi(\beta, \gamma | z, \theta_0).
\end{equation}
Specification of the individual distributions in \eqref{eq:final_prior_form_new} is what remains to fully define the DBSSM.  The priors $\pi(\iota), \pi(\lambda ),$ and $ \pi(\theta_0)$ remain unchanged and can be found in the original paper.  In Section \ref{sec:modeling_framework}, the priors on $\pi(z | \theta_0)$ and $\pi(\beta, \gamma | z, \theta_0)$ are described and updated, when necessary, with the theory developed in this paper.

\end{appendix}

\end{document}